\documentclass[preprint2]{aastex}
\usepackage{psfig}
\usepackage{lscape}
\usepackage{graphics}

\shorttitle{Exoplanet imaging with a PIAAC}
\shortauthors{Pluzhnik et al.}

\begin{document}

\title{Exoplanet Imaging with a Phase-induced Amplitude Apodization
Coronagraph III. Hybrid Approach: Optical Design and Diffraction Analysis}

\author{Eugene~A.~Pluzhnik \altaffilmark{1,6}, Olivier~Guyon \altaffilmark{1},
Stephen~T.~Ridgway \altaffilmark{2},
Frantz~Martinache \altaffilmark{3},
Robert~A.~Woodruff \altaffilmark{4}, Celia Blain \altaffilmark{1} and
Raphael~Galicher \altaffilmark{5}}
\altaffiltext{1}{Subaru Telescope, National Astronomical Observatory of Japan,
650 North A'ohoku Place, Hilo, HI 96720, USA}
\altaffiltext{2}{National Optical Astronomical Observatories}
\altaffiltext{3}{Observatoire de Haute Provence}
\altaffiltext{4}{Lockheed Martin Space Corporation}
\altaffiltext{5}{Ecole Normale Sup\'erieure, Paris, France}
\altaffiltext{6}{Institute of Astronomy of Kharkov National University, Sumskaya
	   35, 61022 Kharkov, Ukraine}

\begin{abstract}
Properly apodized pupils can deliver point spread functions (PSFs) free of
Airy rings, and are suitable for high dynamical range imaging of extrasolar
terrestrial planets (ETPs). To reach this goal, classical pupil
apodization (CPA)
unfortunately
requires most of the light gathered by the telescope to be absorbed,
resulting in poor throughput and low angular resolution. Phase-induced
amplitude
apodization (PIAA) of the telescope pupil \cite{Guyon2003} combines the
advantages of classical pupil apodization (particularly low sensitivity to
low order aberrations) with full throughput, no loss of angular resolution
and little
chromaticity, which makes it, theoretically, an extremely attractive
coronagraph for direct imaging of ETPs.
The two most challenging aspects of this technique are (1) the difficulty to
polish the required optics shapes and (2) diffraction propagation effects
which, because of their chromaticity, can decrease the spectral bandwidth
of the coronagraph. We show that a properly designed hybrid system
combining classical apodization with the PIAA technique can solve both
problems simultaneously. For such a system, the optics shapes can be well
within today's optics manufacturing capabilities, and the  $10^{-10}$
PSF contrast at $\approx 1.5 \lambda/D$ required for efficient imaging of
ETPs can be maintained over the whole visible spectrum. This updated design
of the PIAA coronagraph  maintains the high performance of the earlier design, since
only a small part of the light is lost in the classical apodizer(s).
\end{abstract}

\keywords{direct exoplanet imaging, coronagraphy, apodization, pupil
remapping, diffraction propagation, hybrid optical design}
\maketitle

\section{Introduction}
\label{sec1}

An optical system capable of extremely high contrast imaging
(about $10^{-10}$) at separations  comparable to the telescope's
diffraction limit is critical for direct imaging of extrasolar terrestrial
planets.

Properly apodized telescope pupils (Nisenson \& Papaliolios 2001;
Kasdin et al. 2003), or
designs derived from the classical Lyot coronagraph
\cite{Soummer2003,Kuchner2005} provide the appropriate contrast level.
Unfortunately, they suffer from low throughput, ranging from 0.1 to 0.3,
and large inner working angles (IWAs), above $3\lambda/D$. More efficient
concepts, capable of near 100\% throughput and $\approx\lambda/D$~IWA
exist \cite{Roddier1997,Baudoz2000,Rouan2000}. They however exhibit a
reduced performance for off-axis rays, sufficiently strong to prevent high
contrast on nearby partially resolved stars.

A recently proposed alternative to the ``classical'' pupil apodization
(refered to as  CPA in this work) is to geometrically remap the entrance
pupil of~~the~~telescope~~into~~an~~apodized~~pupil (Guyon

\begin{landscape}
\begin {figure}[p]
\psfig{figure=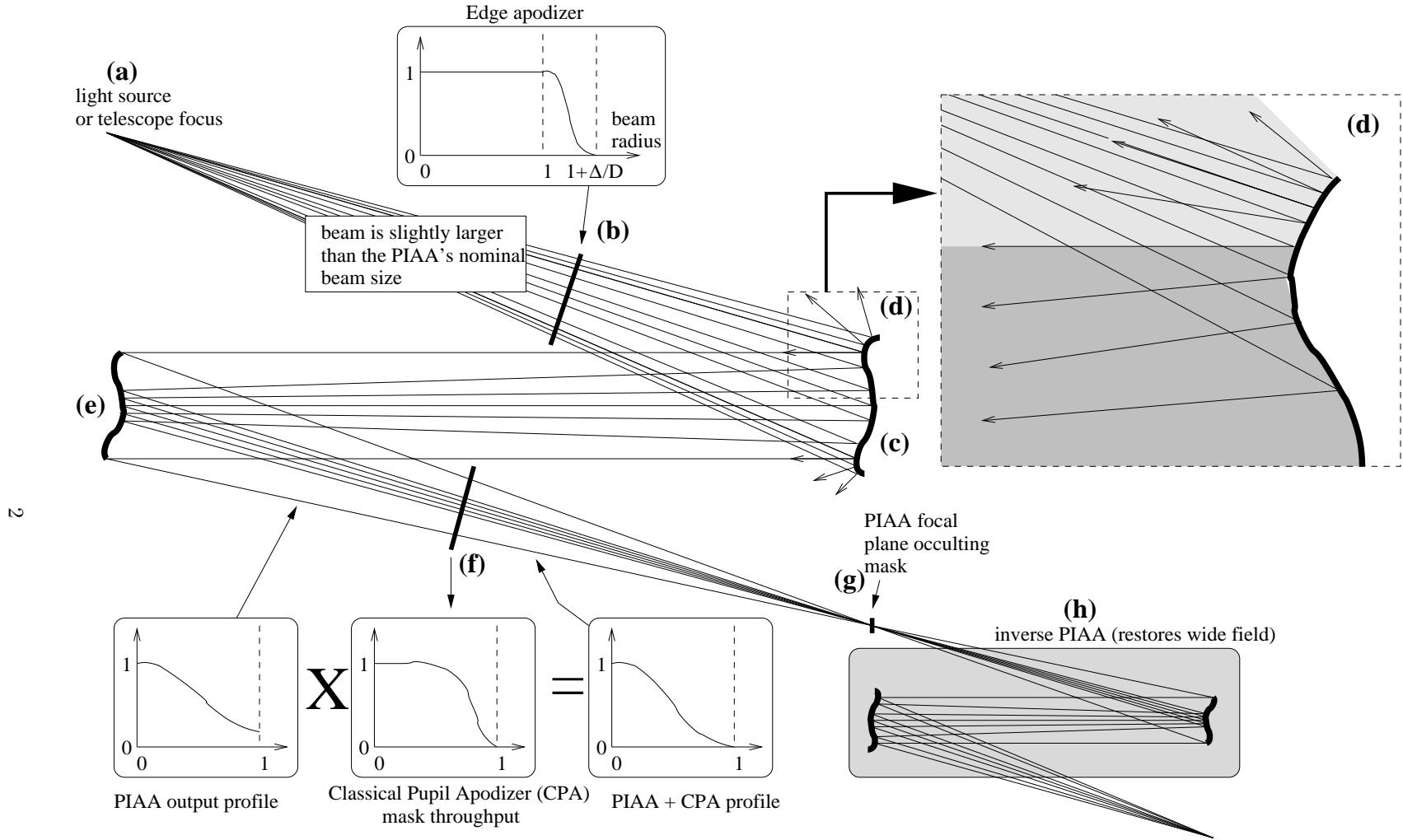,width=1.3\textwidth}
\caption{
Optical layout of a PIAA/CPA hybrid Coronagraph. The system is shown with a
focal
plane input (a) and output (g), but could also be designed to accept and
deliver a collimated beam.
Most of the apodization is performed by the 2 aspheric mirrors M1 (c) and M2
(e), which remap the incoming beam into a truncated gaussian-like profile. A
second
apodization, produced by the classical apodizer (f), removes some of the
light
in the wings of this profile to produce a spheroidal prolate profile. The
opaque focal plane mask (g) efficiently removes the light of the central
source,
while the rest of the field is fed to a PIAA unit mounted backwards (h) to
restore a clean off-axis PSF over a ``wide'' field.
In order to minimize unwanted diffraction effects, the apodization profile
delivered by the aspheric mirrors is carefully chosen to avoid strong
curvature on the M1
mirror. Further mitigation of diffraction effects is obtained by slightly
oversizing the entrance beam and apodizing its outer edge (b). Thanks to a
constant-curvature extension (d) of the first PIAA mirror, this oversized
edge-apodized beam is projected on the second PIAA mirror (e) which therefore
acts as the pupil stop in the system.}
\label{fig1}
\end{figure}
\end{landscape}

\noindent 2003) (this
technique is referred to as PIAA, or phase-induced amplitude apodization,
in this work). This can be done with two aspheric optics, preferably mirrors:
the first aspheric mirror is mostly used to project on the second mirror the
desired beam profile, and the second mirror recollimates (or refocuses) the
output beam. Mirror shapes can easily be computed by solving a differential
equation \cite{Guyon2003,Traub2003}. Although such a \hbox{system}~~
corrupts~~~the~~~telescope~~~isoplanaticity \\
\hbox{(the~~unabberated~~field~~of~~view~~is~~only~~about~~} of $5\lambda/D$
\footnote{Only the sky related angular scale $\lambda/D$, measured for
the principal ray of the system \cite{Guyon2005}, is
used in this paper.}), a wide field of view can be restored by
using the second set of post-coronagraphic PIAA optics \cite{Guyon2003}
which does not affect the coronagraphic performance.

The  PIAA technique combines
many of the advantages found separately in other coronagraphs:
\begin{enumerate}
\item{Very high throughput for the planet's light (nearly 100\%)}
\item{Small inner working angle (slightly larger than $\lambda/D$).}
\item{Excellent achromaticity if implemented with mirrors
(in the geometrical optics approximation).}
\item{Relative insensitivity to pointing errors.}
\end{enumerate}
These advantages have been quantified in several studies of the PIAA
(Guyon 2003; Traub \& Vanderbei 2003; Vanderbei \& Traub 2005).
A detailed analysis of a
complete PIAA coronagraph (PIAAC) design was recently performed
\cite{Guyon2005}, and the performance of the same design was evaluated
for an imaging survey of ETPs with a space telescope \cite{Martinache2005}.
This last study showed that the PIAAC is significantly more efficient
than CPAs. A laboratory experiment,
performed with lenses, has demonstrated beam apodization and imaging with a
PIAA unit \cite{Galicher2005}.

While these studies showed that the PIAAC is, in theory, very efficient for
direct imaging of ETPs, two serious concerns remain unanswered:
\begin{itemize}
\item{{\bf Optics manufacturability.} In the original PIAA design
\cite{Guyon2003}, the outer edge of the first PIAA mirror is highly curved.
This feature, which is essential to obtain the desired apodization, is
extremely difficult to polish.}
\item{{\bf Effects of diffraction propagation.} PIAA units have so far been
designed and studied with geometric and Fraunhoffer approximations. As
recently shown by Vanderbei (2005), differences between diffraction
propagation and geometric/Fraunhoffer optics are not negligible at the
$10^{-10}$ contrast level.}
\end{itemize}

In this work, we will address both issues through the study of a PIAA/CPA
hybrid system. This new design combines a PIAA unit with a mild ``classical''
apodization of the beam.
In \S\ref{sec2}, we focus on the optics manufacturability issue, present our
hybrid design and explain how it solves this challenge.

In \S\ref{sec3} we introduce the diffraction propagation problem
in the PIAA apodizer and describe our method of diffraction calculation.
The effects of diffraction propagation
on the PSF contrast in a poorly designed system and possible
solutions of the problem
are analyzed in \S\ref{sec4}.
Lessons learned from \S\ref{sec4}  are used to
design a much superior
hybrid system which is studied in \S\ref{sec5}.
We give there a broader analysis of the PIAA design tradeoffs
and quantify the performance of such systems for direct imaging of ETPs.


\section{PIAA systems optical designs according to geometrical optics}
\label{sec2}
\subsection{PIAA unit design}
\label{sec2_1}
In its original design, the PIAA optics consist of two aspherical mirrors
M1 and M2 (Fig.~\ref{fig1}).
In the focus-to-focus system studied in this paper, the source is
``collimated'' by the first mirror M1 and reimaged by the second mirror M2.

The remapping function $f(r_1)$ is determined in a such way that
the total flux within the radius $r_1$ of the input beam is equal to
the total flux within the radius $f(r_1)$ of the output beam.
For any desired remapping function $r_2=f(r_1)$, where  $r_1$ and $r_2$
are the geometrical radii on mirrors M1
and~~M2~~where~~a~~ray~~emitted~~by~~an~~on-axis
\begin{landscape}
\begin {figure}
\psfig{figure=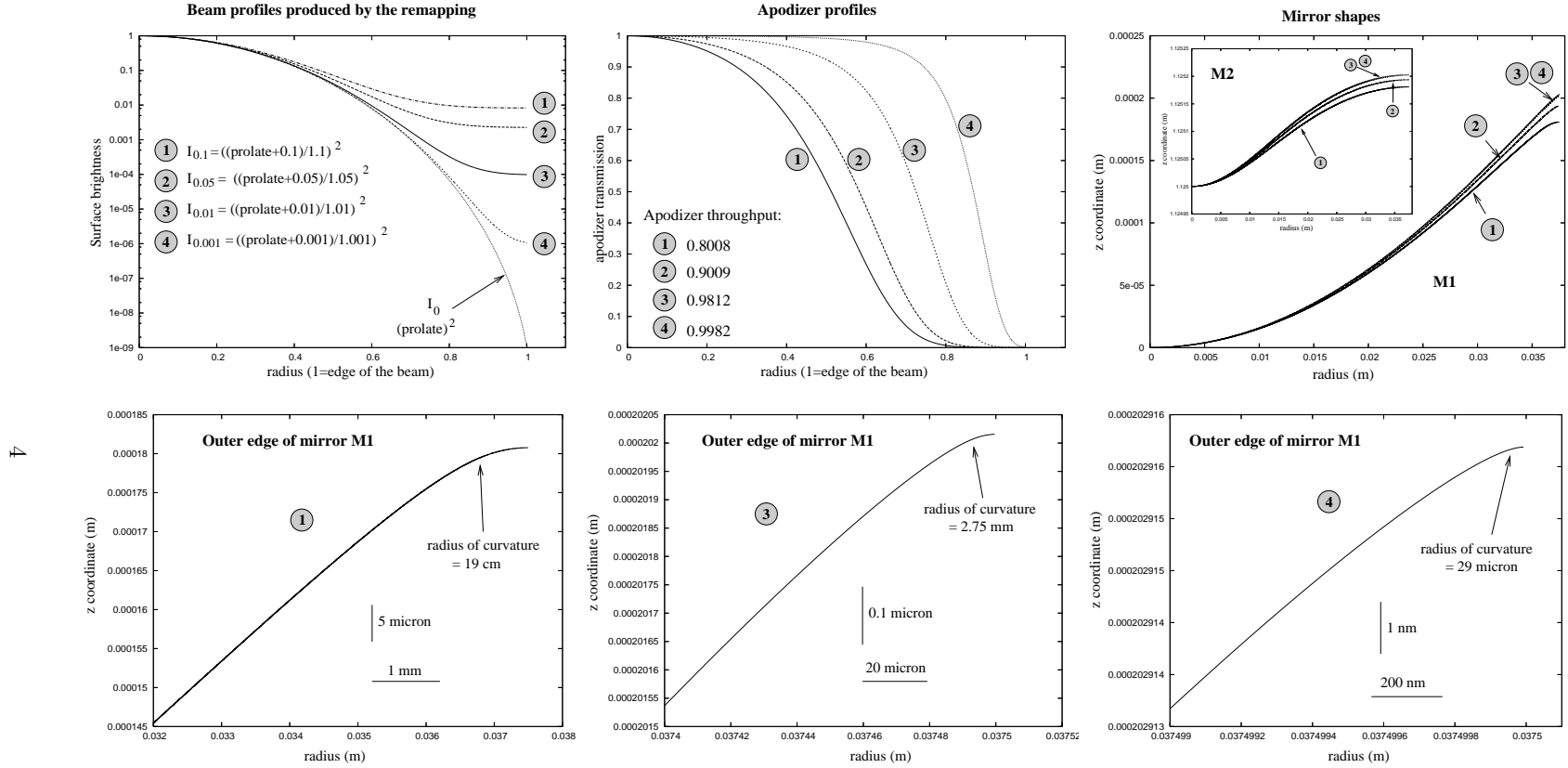,width=1.3\textwidth}
\caption{Surface brightness profiles $I_{\alpha}$ for four
values of $\alpha$ (top left), and the prolate profile $I_0$. For each
profile, the apodizer profile required to obtain the desired prolate
beam is shown for an apodizer placed after the PIAA unit (top center).
The aspheric terms in the mirror shapes are shown (top right) for M1 and
M2. In the bottom panels, the shape of M1's edge is shown in more detail
for three of the four values of $\alpha$.}
\label{fig2}
\end{figure}
\end{landscape}
\noindent source is reflected, the shape of the mirrors
is obtained by solving the differential equation \cite{Guyon2003}
\begin{equation}
\frac{dM_1}{r_1}=\frac{dM_2}{r_2}=\sqrt{1+\left(\frac{M_2-M_1}{r_1-r_2}\right)^2}-
\frac{M_2-M_1}{r_1-r_2},
\label{eq1}
\end{equation}
and can be written as
\begin{equation}
M_i(r_i)=BP_i(r_i)+P_i(r_i).
\label{eq2}
\end{equation}
In Eq.~\ref{eq2} $BP_i(r_i)$ is a paraboloid of rotation describing
the base of the i-th mirror and $P_i(r_i)$ describes the modification
of the base shape to produce the desired apodization profile. The desired
apodization profile is formed mainly by the first
mirror M1, while the mirror M2 is mostly used to correct phase errors
produced by the first mirror.

This first set of optics creates a properly apodized pupil beam which is
used to form a high-contrast image. An occulting spot can then block the
light of the central source. Off-axis aberrations introduced by the PIAA
can be corrected by a second PIAA unit. The role of this second unit
is only to ``sharpen'' the images of off-axis sources and
it is therefore not discussed in this paper.

\subsection{Optics shapes: the case for a hybrid PIAA/CPA approach}
\label{sec2_2}

A prolate spheroidal beam amplitude profile can be shown to offer a
contrast exceeding $10^{-10}$ over a 360 degree search angle
(Soummer et. al 2003).
We denote $I_0(r)$ its surface brightness profile.
Unfortunately, this beam profile cannot realistically be obtained directly
by remapping of the entrance pupil: the very faint outer edge of the prolate
function (the profile edge to center brightness ratio is about of
$10^{-9}$) would
require a ``dilution''
of the incombing beam's edge by a factor approximately $10^8$: the outer 1\%
(in radius) of the apodized beam is to contain as much light as the outer
$10^{-8}$\% of the unapodized beam. The optics shapes required to perform
this task exhibit a narrow highly curved edge on the first mirror, which is
both extremely challenging to manufacture and essential to reach the desired
apodization in one step.

At least two solutions exist to mitigate this problem: splitting the
remapping into several steps (this requires additional aspheric optical elements)
or sharing the apodization between a remapping system and a ``classical''
apodizer. This second option seems at present less costly and is adopted in
this work.
In this scheme, the manufacturability of the optics must be balanced against
the overall throughput of the apodizer. Figure \ref{fig2}
illustrates four possible combinations, obtained by designing the pupil
remapping unit to deliver a beam profile obtained by adding a constant
$\alpha$ to the prolate function:
\begin{equation}
\label{eq3}
I_{\alpha}(r) = \left(\frac{\sqrt{I_0(r)}+\alpha}{1.0+\alpha}\right)^2. 
\end{equation}
The corresponding apodizer mask, required to produce the final beam profile
$I_0$, is
\begin{equation}
T(r) = I_0(r)/I_{\alpha}(r)
\label{eq4}
\end{equation}
and is also shown in Fig.~\ref{fig2}. For higher values of
$\alpha$, the throughput of the system is lower, but the optics become easier
to manufacture (the outer edge of M1 becomes more gentle).
Mirror shapes shown in Fig.~\ref{fig2} assume a 75~mm beam
diameter on both M1 and M2 mirrors
(refered to as the {\bf working beam diameter}), a 1.125~m separation
between M1 and M2, and a
collimated input/ouput beam.
With $\alpha=0.1$, the system throughput
is only 80\%, but the M1 mirror shape is very ``friendly'', with a 19cm
minimum radius of curvature in the $\approx 2$mm wide outer edge. In a
$\alpha=0.01$ system, only 2\% of the light is lost in the classical apodizer,
and the
optics shape, although much more challenging, appears to be manufacturable
(2.75mm minimum radius of curvature over the last $\approx$20 $\mu$m of M1).
With $\alpha=0.001$, the throughput is excellent (99.8\%) but M1 appears to
be extremely difficult to manufacture: a well-controlled bend with a 29 $\mu$m
curvature radius over the last $\approx 200~nm$ of M1 would be required.
Systems designed to require very little absorption by the apodizer are
therefore very difficult to manufacture.
We have not attempted to perform accurate geometrical and diffractive
simulations for
small values of $\alpha$, as the required number of sampling points for a
such an  analysis would be prohibitively high (for example, accurate
simulation of a
$\alpha = 0.001$ system would require a $\approx 5~nm$ sampling at the outer
edge of M1). For these reasons, our study is limited to systems that are both
manufacturable and easy to simulate: systems for which the apodizer removes
at least 1\% of the flux.

\begin {figure*}
\psfig{figure=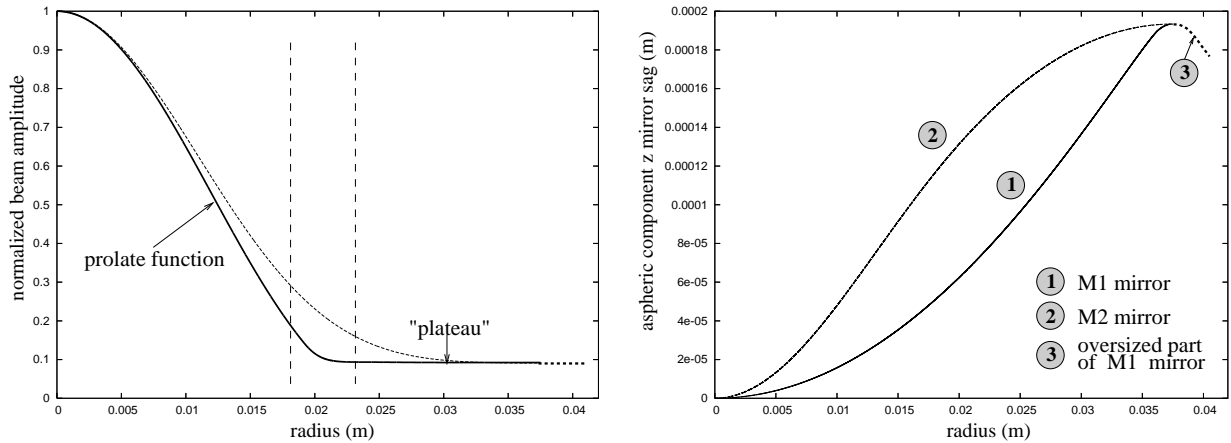,width=1.0\textwidth}
\caption{
Shape of M1 and M2 mirrors ({\it right}) to
produce the amplitude Profile I ({\it left, solid})
designed for our laboratory experiment.
Only the aspheric terms in the mirror shapes are shown here.
The two mirrors are separated by 1.125~m in the z direction.
The amplitude profile $\sqrt{I_{0.1}(r)}$ ({\it left, dashed})
is also shown for a comparison.
Oversized parts of
the working beam and M1 mirror are drawn with bold dashed lines.}
\label{fig3}
\end {figure*}

In order to make mirror M1
easier to polish, the minimal surface brightness in the beam produced by the
remapping optics needs to be kept above some level.
The beam profile to be produced by the remapping optics does not have to be
chosen according to Eq. \ref{eq3}.
It is therefore tempting to design a remapping system with an output
intensity profile that
closely follows the $I_0$ profile within the central region of the beam with
a relatively quick transition to a ``plateau'' in the outer part of the beam.
For the same ``plateau'' level, this new profile offers a throughput
higher than the $I_{\alpha}$ profile.
An example of such a profile (Profile I)
is shown
in Fig.\ref{fig3}~.
This profile has been chosen for our
upcoming laboratory experiment.
In this paper  it is used to demonstrate some
diffraction effects encountered in a PIAA system.
The central part of this profile is a prolate function
continued with a constant level near  the edge of the pupil.

Our hybrid PIAA/CPA apodization requires the use of an apodizing mask, just
as a ``classical'' apodized pupil coronagraph does. Pupil apodizers for
coronagraphs are technologically difficult to manufacture: the optical
density needs to be well controlled and achromatic.
Fortunately, the tolerances for the
apodizer are easier to meet in the hybrid PIAA/CPA coronagraph than for a CPA
coronagraph:
\begin{itemize}
\item{In the PIAA/CPA design, the pre-apodizer beam is already apodized, and
the required maximal apodizer's optical density is lower than if it were used
without the PIAA.}
\item{The apodizer only affects  regions of the beam where the surface
brightness is low. An error in optical density has therefore  a smaller
effect on the PSF contrast than if the apodizer were used by itself. In the
central region of the beam the apodizer's transmission is almost constant,
and close to 100\%. Large variations in the apodizer's transmission only
occur in the fainter outer parts of the remapped beam.}
\end{itemize}

\section{Diffraction propagation in a PIAA/CPA hybrid system}
\label{sec3}
\subsection{Diffraction effects and PSF contrast chromaticity}
Diffraction effects are most strongly introduced by discontinuities or sharp
transitions. In a PIAA system, the outer edge of the beam (which could be
defined by the edge of mirror M1) and the ``sharp'' bend near M1's boundary
are therefore of particular concern. While we
will closely examine these effects and propose solutions to mitigate them
in the following sections, we discuss here briefly their impact on the
coronagraph performance.

Diffraction effects can modify the coronagraph behaviour, such that the
coronagraph output is different from what geometrical optics theory
predicts \cite{Vanderbei2005a}: the PIAA unit, as presented in
\S\ref{sec2_1},
relies on geometrical optics to apodize the beam. As will be illustrated
in the following sections, the difference between the expected
(from geometrical optics) and actual (taking into account diffractive
effects) apodized beams is quite small in most cases: less
than $\lambda/100$ RMS in phase in the visible. This difference is most
likely not noticeable in practice, for two reasons:
\begin{itemize}
\item{At this level of accuracy, a coronagraph optical system is relying on
fine wavefront control rather than the intrinsic figure of the optics.}
\item{The PIAA optics are manufactured as a set, one serving as the null
for the other one. The diffraction effects therefore naturally occur during
the testing of the optics: polishing to the null will compensate for the
diffraction effects.}
\end{itemize}
Even if the diffraction effects are significant, they can easily be
integrated within the design of optical elements: residual phase errors
can be projected on either M1 or M2; amplitude errors can also be
cancelled by slight modification of M1's shape.

The first step of this process   is to design the mirror
shapes geometrically \cite{Guyon2003} to obtain the desired amplitude
profile $A_g(r)$.
The geometrically designed M1 mirror
produces a diffraction pattern on the M2 surface with
the OPD which
is slightly different from the one predicted by geometrical optics laws.
This difference can be corrected by changing of the M2 shape.
For small incidence angles, the appropriate shape change is approximately
$(OPD_g-OPD_d)/2$ in z coordinate. We denote  $A_d(r)$ the output beam
amplitude profile obtained by this OPD-corrected system.
The resulting amplitude residuals can be then compensated by changing both
M1 and M2 mirror shapes to geometrically produce an output beam with
amplitude equal to $A_g^2(r)/A_d(r)$. A few iterations of this process
rapidly converge
to the solution  for the mirror shapes with $A_d(r)=A_g(r)$ when the
diffraction effects are not very large.
We have successfully used
a similar algorithm to design an off-axis focus-to-focus PIAA optical system, in
which the optics shapes are non trivial, but can be obtained iteratively by
correcting M1 and M2's shapes to cancel residual beam aberrations. In an
hybrid PIAA/CPA system, amplitude aberrations can also be cancelled by the
apodizer at a small cost in throughput.

A far more serious concern is the chromaticity of the diffraction effects:
careful design of a PIAA/CPA system and/or fine correction by deformable
mirror(s) can only cancel diffraction effects at a single wavelength. Strong
diffraction effects therefore limit the spectral bandwidth over which the
system can maintain an appropriate contrast. This is a general problem in
high-contrast coronagraphy: mirror edges, mask edges and amplitude/phase
aberration on optical elements all introduce wavelength-dependant effects
through diffraction. For example, a pure optical pathlength difference (OPD)
aberration will propagate (through diffraction propagation) into a chromatic
OPD and amplitude aberration. While solutions to mitigate these problems in
classical optical designs are relatively well known (oversizing optics which
are not in a pupil plane, minimizing aberrations introduced by optics,
conjugating DMs  to the source of aberrations, etc...),
it is largely unknown to what degree diffraction effects affect the PIAA
system, and how to mitigate them.

The goal of this work is therefore to quantify the effect of diffraction
propagation on the PSF contrast chromaticity  and to identify solutions
to mitigate it. Ultimately, we wish to verify if the PIAA coronagraph is
truly achromatic, as geometrical optics predicts. All simulations presented
in this work are for a PIAA/CPA hybrid design, which, unlike a pure PIAA
system, seems manufacturable (see \S\ref{sec2_2}). In each
optical configuration, it is assumed that the system is
diffraction-compensated for $\lambda_0 = 0.633\mu$m: the PIAA unit optics
shapes (and/or the DM shape) are such that no phase aberration is present
at $\lambda_0$; the classical apodizer is also designed such that no
amplitude aberration is present at $\lambda_0$. We also assume that the PIAA
optics shapes (and/or the DM) are achromatic: the OPD is independant of
$\lambda$; likewise, the throughput of the classical apodizer is assumed to
be achromatic.
\subsection{Computation of the Rayleigh -- \\Sommerfeld integral}

\begin {figure*}
\hspace{7ex}
\psfig{figure=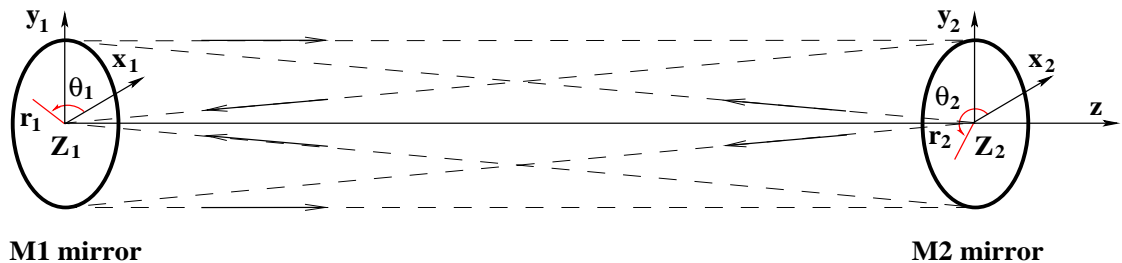,width=0.9\textwidth}
\caption{Geometry of the diffraction problem:
A point source at Z2 in the center of mirror M2 is collimated by
mirror M1, and then reimaged by M2 to Z1 at the center of M1.}
\label{fig4}
\end{figure*}

For computational efficiency only symmetric, on-axis systems are studied in
this paper.
The geometry adopted for our diffraction computations is shown in Fig.~\ref{fig4}.
The z axis of the coordinate system passes through the centers of the
mirrors M1 and M2. The centers Z1 and Z2 of M1 and M2 mirrors are at
$z_1=0$ mm and $z_2=1125$ mm respectively. A point source is placed in the
center Z2 of M2 mirror,
while its image is formed in the center Z1 of M1 mirror
(focus-to-focus system). Polar coordinates
$(r_1,\theta_1)$ and $(r_2,\theta_2)$ are used to describe
respectively M1 and M2 surfaces. The point source emits a spherical
monochromatic wave which is reflected and diffracted by  M1.
The diffracted wave is focused by M2.

\begin {figure*}
\psfig{figure=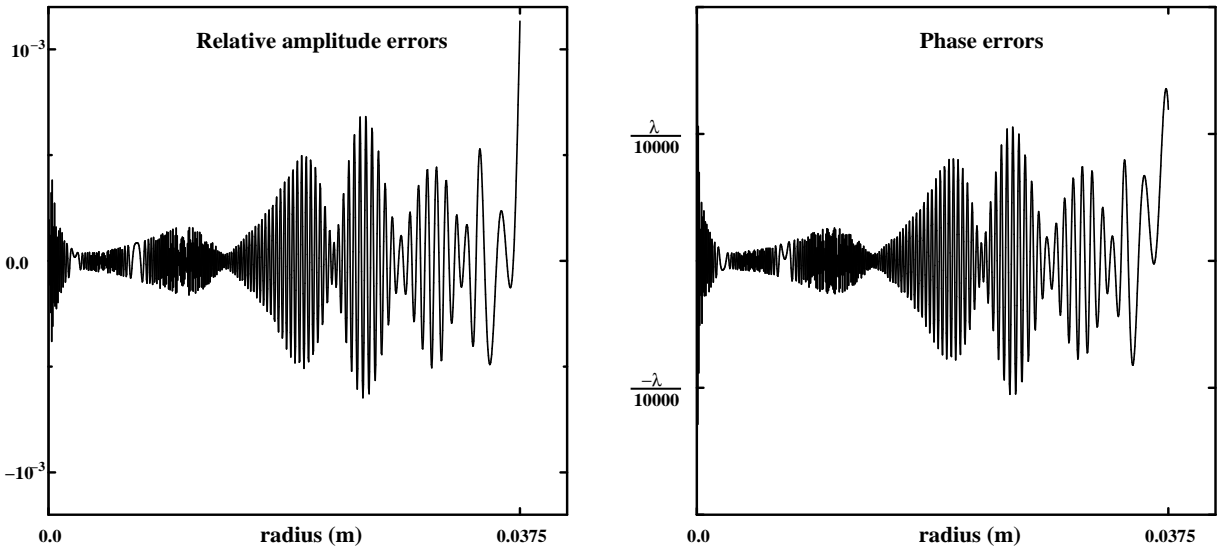,width=1.0\textwidth}
\caption{ Relative amplitude difference and phase difference  on the M2
mirror surface between diffraction calculation results obtained
with two samplings, differing by a factor of 4.}
\label{fig5}
\end{figure*}

The full Rayleigh-Sommerfeld diffraction integral for propagation
between two surfaces M1 and M2 is given by:
\begin{eqnarray}
U({\bf r_2})=\frac{1}{\lambda}\int\int dx_1dy_1 A({\bf r_1})
e^{i\varphi({\bf r_1})}\biggl[
\frac{1}{k
l}-i\biggr]\times \nonumber\\
\frac{M_2-M_1}{l}\frac{\exp(ikl)}{l},\mbox{\phantom{aaaaaaaaaaaaaaaa}}
\label{eq5}
\end{eqnarray}
where $k=2\pi/\lambda$, $l=|{\bf r_2}-{\bf r_1}|$, and the point source
emits a spherical monochromatic wave with amplitude
$A({\bf r_1})=A_0/\sqrt{r_1^2+(z_2-M_1)^2}$ and
phase $\varphi({\bf r_1})=2\pi\sqrt{r_1^2+(z_2-M_1)^2}/\lambda$ on the
surface of M1 mirror. In polar coordinates Eq.~\ref{eq5} can be written as
\begin{eqnarray}
U(r_2)=\frac{2}{\lambda}\int_0^R A(r_1) e^{i\varphi(r_1)} (M_2-M_1) r_1 dr_1 \nonumber\\
\int_0^\pi
\biggl[ \frac{1}{k l}-i\biggr] \frac{\exp(ikl)}{l^2}
d\theta_1, \mbox{\phantom{aaaaaaa}}
\label{eq6}
\end{eqnarray}
where R is the mirror radius.
Unfortunately, the accuracy of well known approximations for
diffraction integrals (such as the Fresnel approximation)
is not sufficient for coronagraphic applications \cite{Vanderbei2005a}.
That is why we
performed direct numerical integration of Eq.~\ref{eq6} to estimate the
diffraction effects in the PIAA system by using a
Fujitsu PrimePower2000 supercomputer at Subaru Telescope
\cite{Ogasawara2004}.
The supercomputer consist of 128 processors with the maximal performance
of about 170 Gflops \cite{Dongarra2002}.
Calculation of the integral in
Eq.~\ref{eq6} was
performed for $\lambda=0.633\mu$m and $\lambda=0.7\mu$m with a sampling
of about $3 R/\lambda$ (the step is equal $2\times 10^{-7}$ m) points
in the radial direction and 10000 points in the angular direction.
This sampling provides us with an accuracy in our diffraction calculation
of about 0.05\% in amplitude and $2\pi/10^4$ radian in phase (Fig~\ref{fig5}).
These accuracy estimates are based on comparison of diffraction calculations
with $3.0 R/\lambda$ radial~$\times$~10000 angular point and
$1.5 R/\lambda$ radial~$\times$~20000 angular point sampling.
The focal plane complex amplitude
distribution was obtained by a discrete
Hankel transform of the complex amplitude distribution on a reference sphere
near the M2 mirror, assuming geometrical propagation between the mirror and
the sphere.

\section{Diffraction effects and possible solutions: examples from a poorly
designed system}
\label{sec4}

In this section, a PIAA/CPA hybrid system which has not been designed to
mitigate diffraction propagation effects is studied. The design adopted,
as will be demonstrated in this section, incorporates several bad choices,
which makes it inadequate for achromatic coronagraphy at the $10^{-10}$
contrast level. It however provides us with a convenient example to
illustrate the different diffraction propagation effects encountered, and
explore solutions. Lessons learned from this exercise will be
used in \S\ref{sec5} to design a much superior PIAA/CPA
system which is largely insensitive to diffraction propagation effects.

\subsection{Presentation of the design and diffraction through the system}
\label{sec4_1}

The PIAA unit studied in this section is built to deliver a beam profile
with a relatively sharp transition between an inner prolate spheroidal
profile and a flat plateau at 0.1 times the peak amplitude
(Fig.~\ref{fig3}). A classical
apodizer, placed downstream of the PIAA unit, then converts this profile
into a pure prolate spheroidal, mostly by attenuating the ``plateau''.
In this system, a hard stop on M1 acts as a pupil stop. The size of this
pupil stop is equal to the useful beam diameter.

\begin {figure*}
\psfig{figure=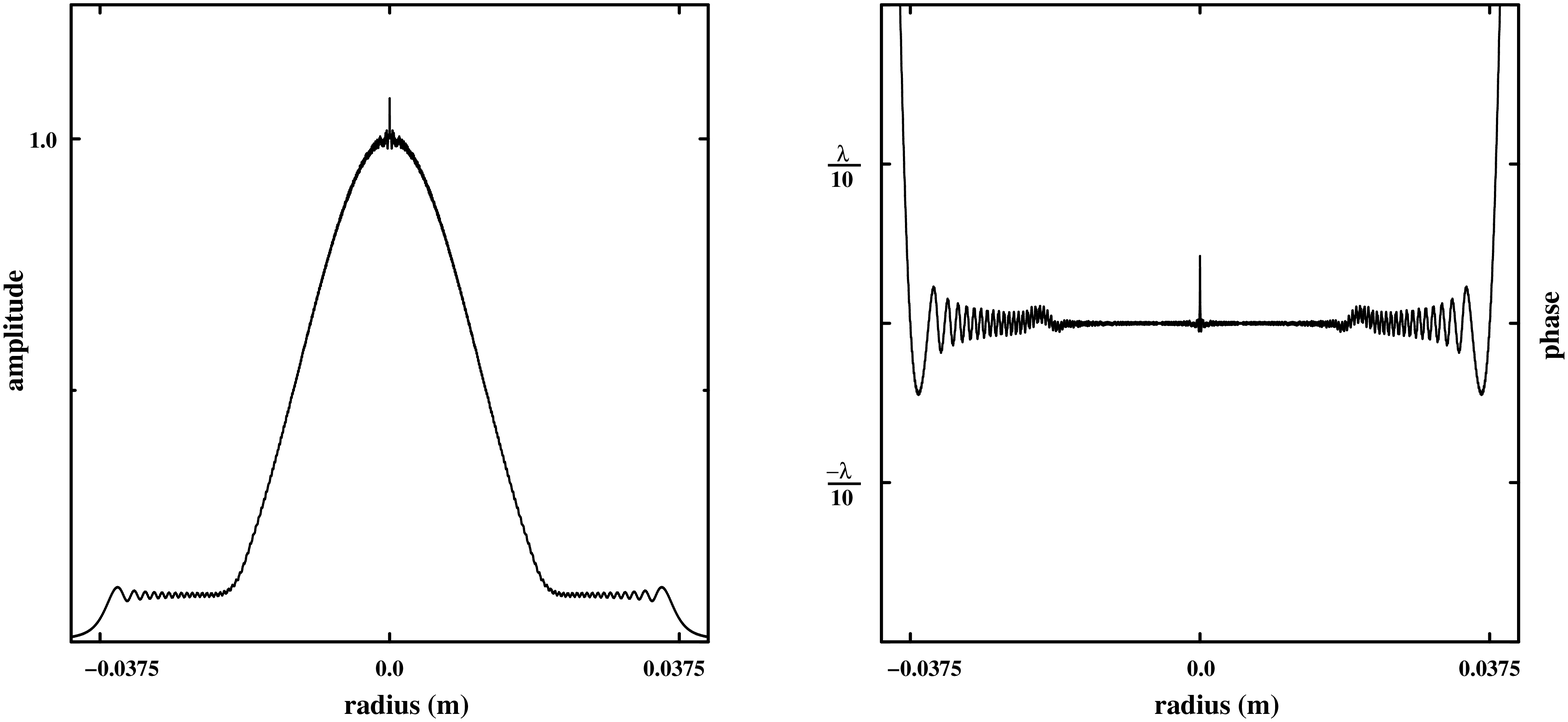,width=1.0\textwidth}
\caption{Diffraction effects in PIAA apodizer, prior to the classical
apodizer. Amplitude and phase distributions on the sufrace of  mirror M2
for the amplitude Profile I are shown here.}
\label{fig6}
\hspace{5ex}
\psfig{figure=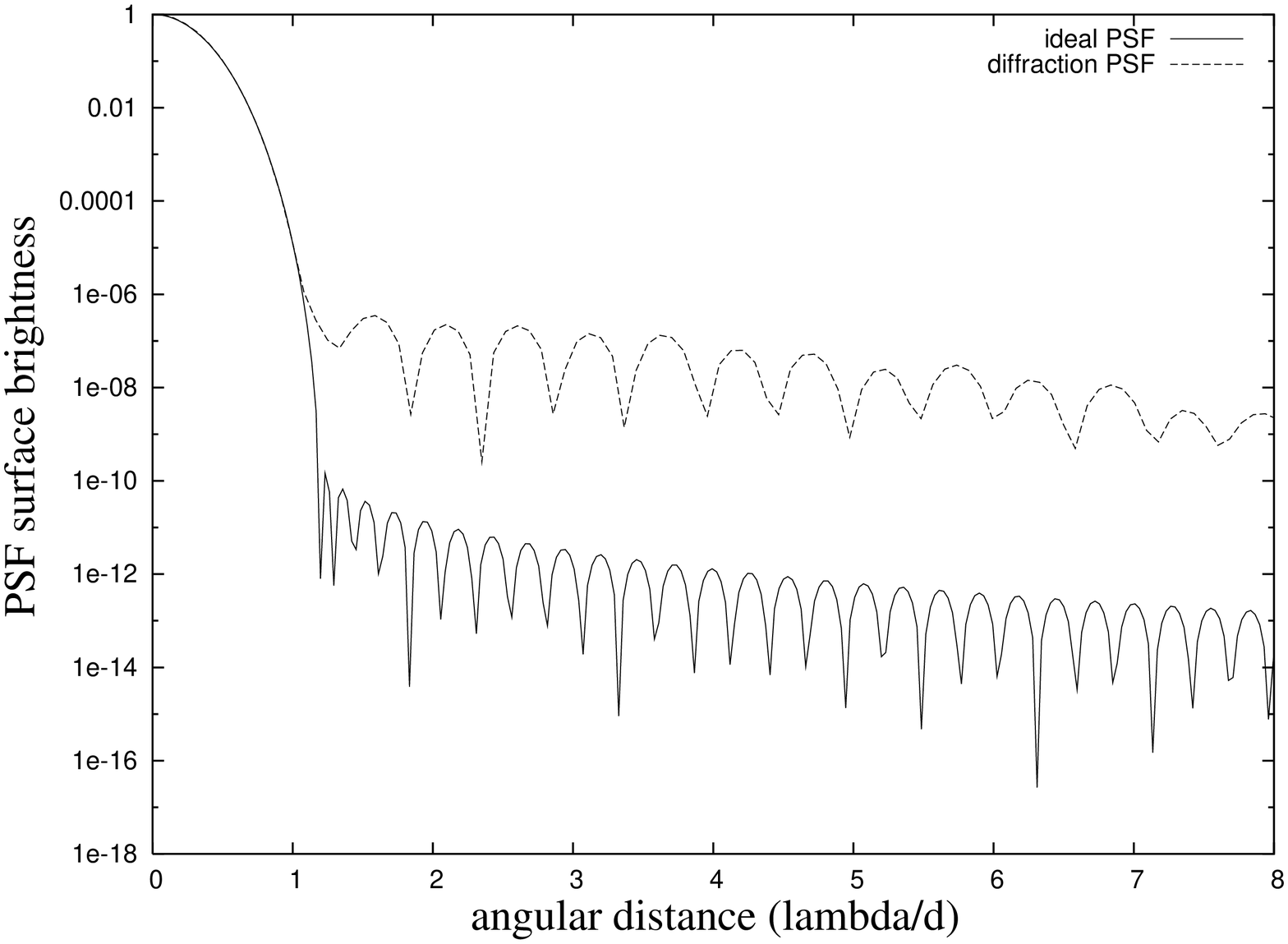,width=0.85\textwidth}
\caption{Diffraction effects in PIAA/CPA hybrid system
based on amplitude Profile I: ideal  and diffraction  PSFs are shown.
The system is designed to deliver
$10^{-10}$ contrast at $1.5\lambda/D$ assuming geometrical optics laws.}
\label{fig7}
\end{figure*}

Two diffraction features can be easily observed in the output beam
(Fig.~\ref{fig6}):
\begin{itemize}
\item{High spatial frequency amplitude and phase oscillations
getting stronger in the outer part of the beam.}
\item{A large peak/hole  in the center of the beam (the so called
Arago spot).}
\end{itemize}
Our diffraction propagation simulation  shows that
the PSF contrast at 2$\lambda/D$ in this system is
$10^{-7}$. This PSF is shown  in Fig.~\ref{fig7}
and differs significantly  from the PSF computed without
diffraction propagation effects.

Because of their high spatial frequency and  wavelength dependence,
it seems to be difficult or impossible to correct
the diffraction effects in a wide bandpass
by using any combination of the PIAA mirror correction,
a classical apodizer and/or
a deformable mirror.
A suitable method to control these diffraction features in the PIAA unit is
considered below.

\subsection{Oversizing and  edge apodization of the entrance beam}
\label{sec4_2}

\begin {figure*}
\hspace{5ex}
\psfig{figure=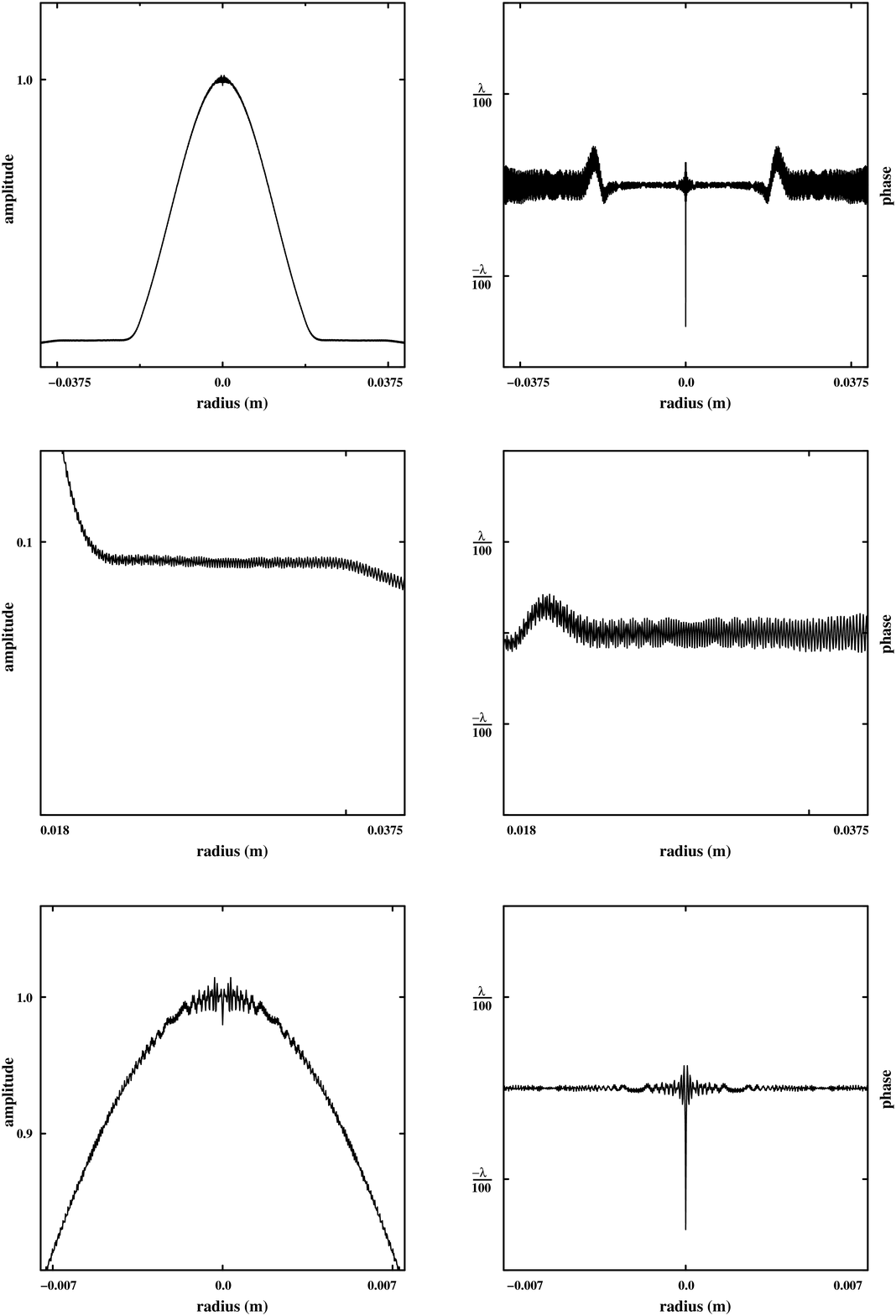,width=0.83\textwidth}
\caption{Diffraction effects in PIAA apodizer with an ``oversized'' beam.
 Amplitude and phase distributions on the sufrace of M2 mirror for amplitude
Profile I
are shown. The core and wings of the amplitude profile are presented separately
in the second and the third rows.}
\label{fig8}
\end{figure*}

It should be noted that the diffraction features identified in
\S\ref{sec4_1} (high frequency oscillations and Arago spot)
arise mainly  from the sharp edge of the input beam
\cite{Rabinowicz1965}.
The period and amplitude
of these oscillations are decreasing  when the distance from the edge
of the beam is increasing.
To reduce this effect, we now consider an M1 mirror with a radius 10\% larger
than the radius of the M2 mirror
(Fig.~\ref{fig1}, feature (d)).
This "oversized" mirror is receiving an equally oversized beam.
The shape of the M1 mirror is designed
to continuously extend the input profile at a constant
amplitude level
outside the working aperture radius (dashed line in Fig.~\ref{fig3}).
Such a mirror still produces
high frequency diffraction oscillations near the boundary of mirror M2
(Fig.~\ref{fig8}),
but they are now about 10 times smaller
than the errors previously seen in the system without this mirror
oversizing.
Unfortunately, these diffraction features are still
highly chromatic and relatively strong.
Two methods to further reduce them can be proposed, namely:
\begin{enumerate}
\item using  a carefully chosen edge apodizer for the M1 mirror;
\item using a destructive phase interference arising from toothed aperture
at the M1 mirror \cite{Shirley1996}.
\end{enumerate}
Both edge apodizing and edge toothing can be performed outside of the
working aperture and have no large effect on the optical design
within the working aperture.

\begin {figure*}
\hspace{5ex}
\psfig{figure=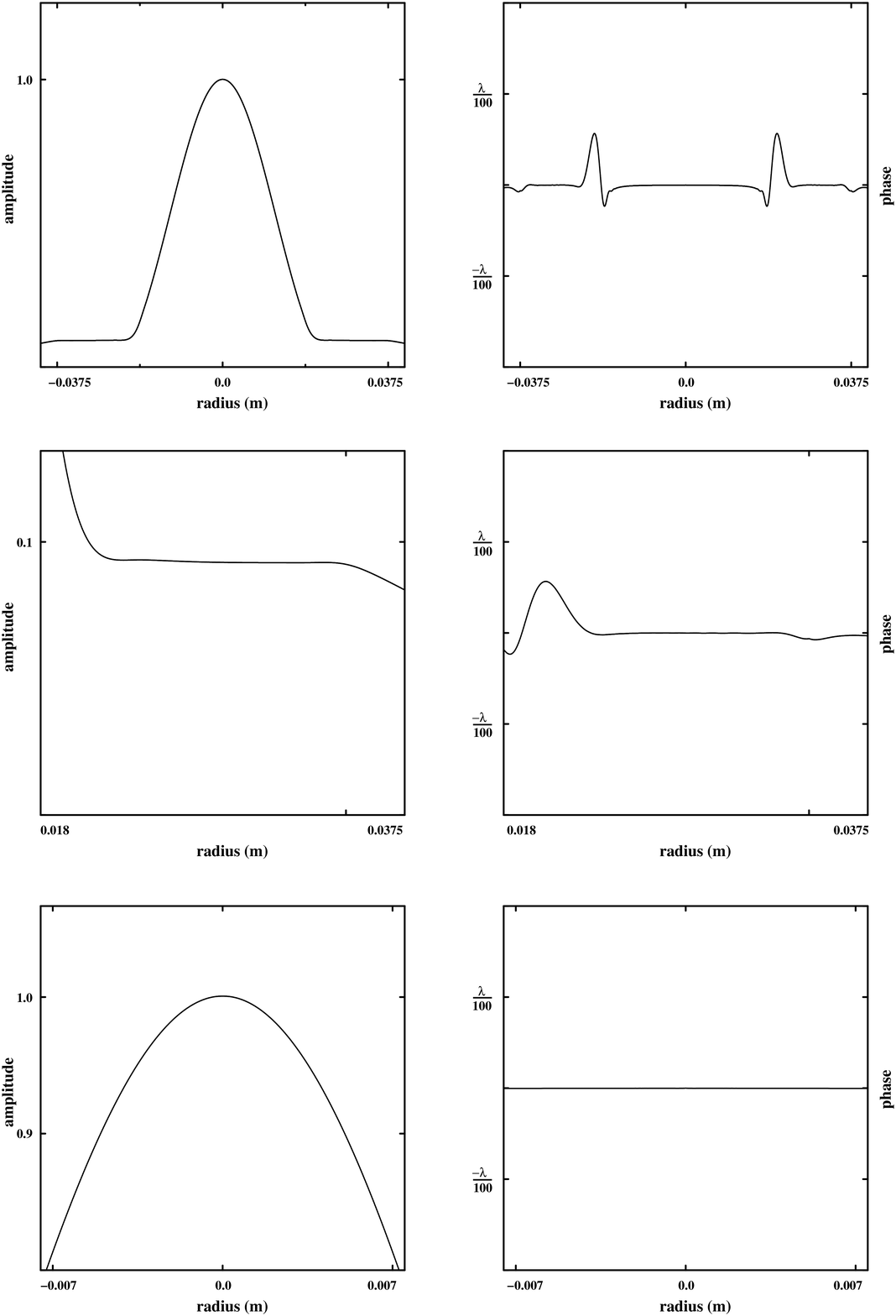,width=0.83\textwidth}
\caption{Diffraction effects in PIAA apodizer with an oversized and
edge-apodized beam. Amplitude and phase distributions on the sufrace of M2
mirror for amplitude Profile I are shown. The core and wings of the amplitude
profile are presented separately in the second and the third rows.}
\label{fig9}
\end{figure*}

We choose here to implement solution (1) with a 10\% cosine taper window to
apodize the sharp edge of the ``oversized'' M1 mirror.
Such an apodization smooths the beam edge within  the
0.0375--0.04125~m radius interval and does not change the complex field
within the working aperture. The diffracted field on the M2 surface
for this "diffraction-free" system (Fig.~\ref{fig9}) does not show either significiant high
frequency oscillations or the Arago effect. The maximal phase error is less
than $\lambda/150$  and rms of phase and amplitude errors are $\lambda/780$
and 0.3\% respectively.

\begin {figure*}[p]
\hspace{5ex}
\psfig{figure=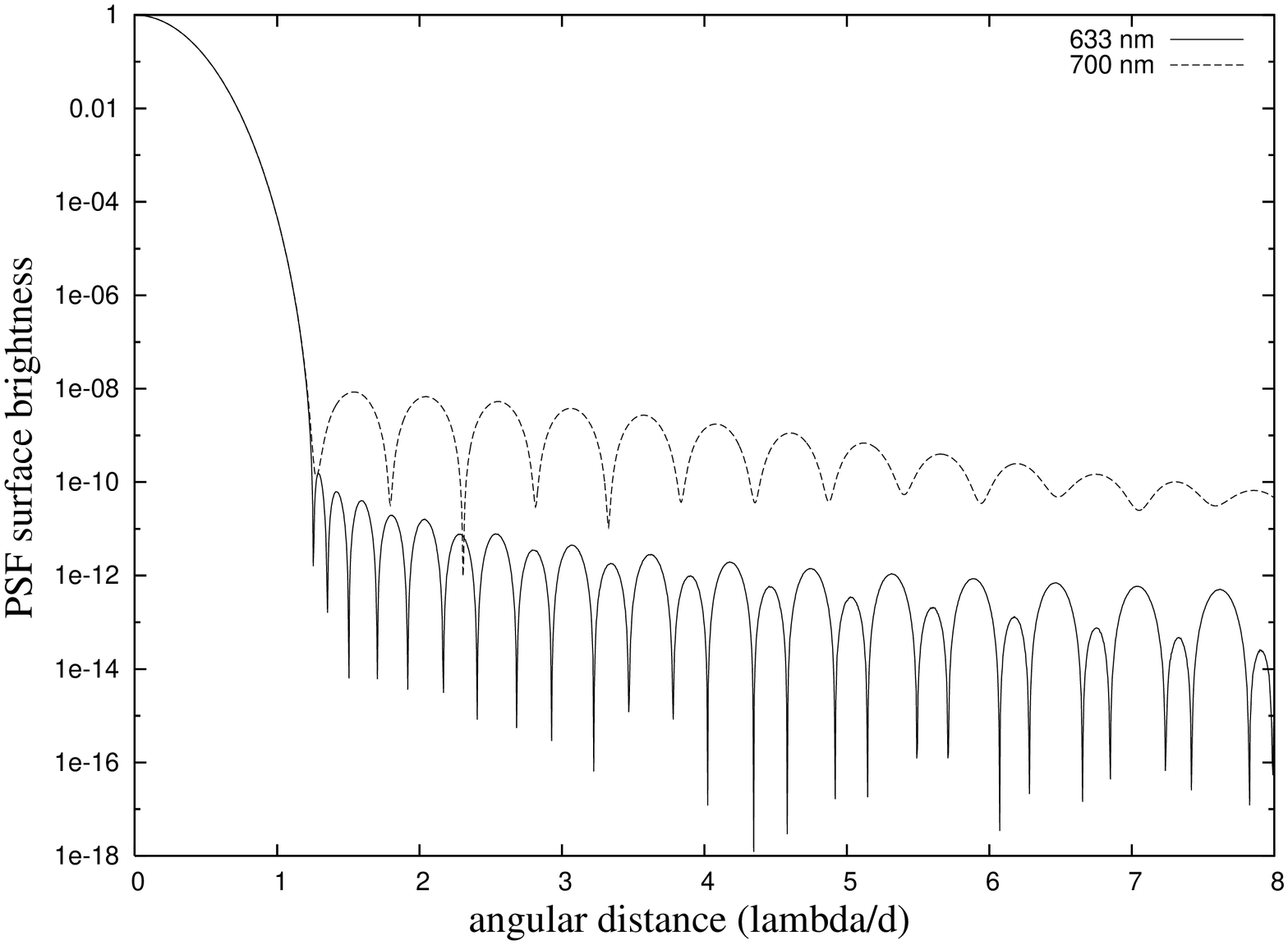,width=0.78\textwidth}
\caption{PSF chromaticity in a "diffraction-free" PIAA/CPA hybrid system
utilizing a poorly designed PIAA unit. The PIAA unit, in this example,
is built to deliver the amplitude Profile I. Both the amplitude profile
and diffraction residuls are corrected to reach a $10^{-10}$ contrast level
at $1.5\lambda/D$ for $\lambda=0.633\mu$m.
}
\label{fig10}
\hspace{5ex}
\psfig{figure=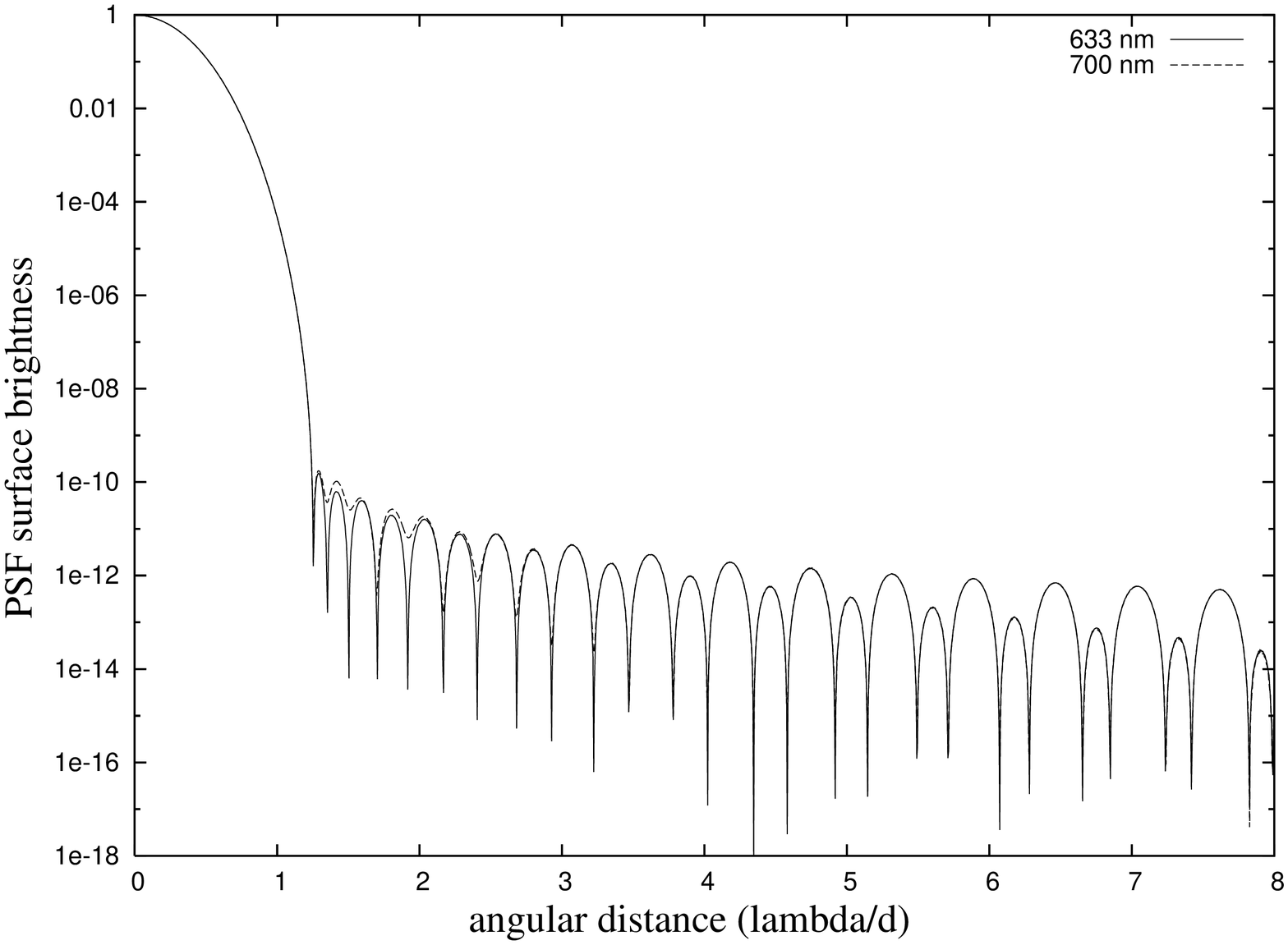,width=0.78\textwidth}
\caption{PSF chromaticity in a "diffraction-free" PIAA/CPA hybrid system
utilizing a properly designed PIAA unit. The PIAA unit by itself, without
the apodizers,
is built to deliver the $I_{0.1}$ beam profile. The PSF contrast is almost
identical at $0.633\mu$m and $0.7\mu$m, suggesting that this design offers a
very high
level of achromaticity.}
\label{fig11}
\end{figure*}

\begin {figure*}[t]
\hspace{5ex}
\psfig{figure=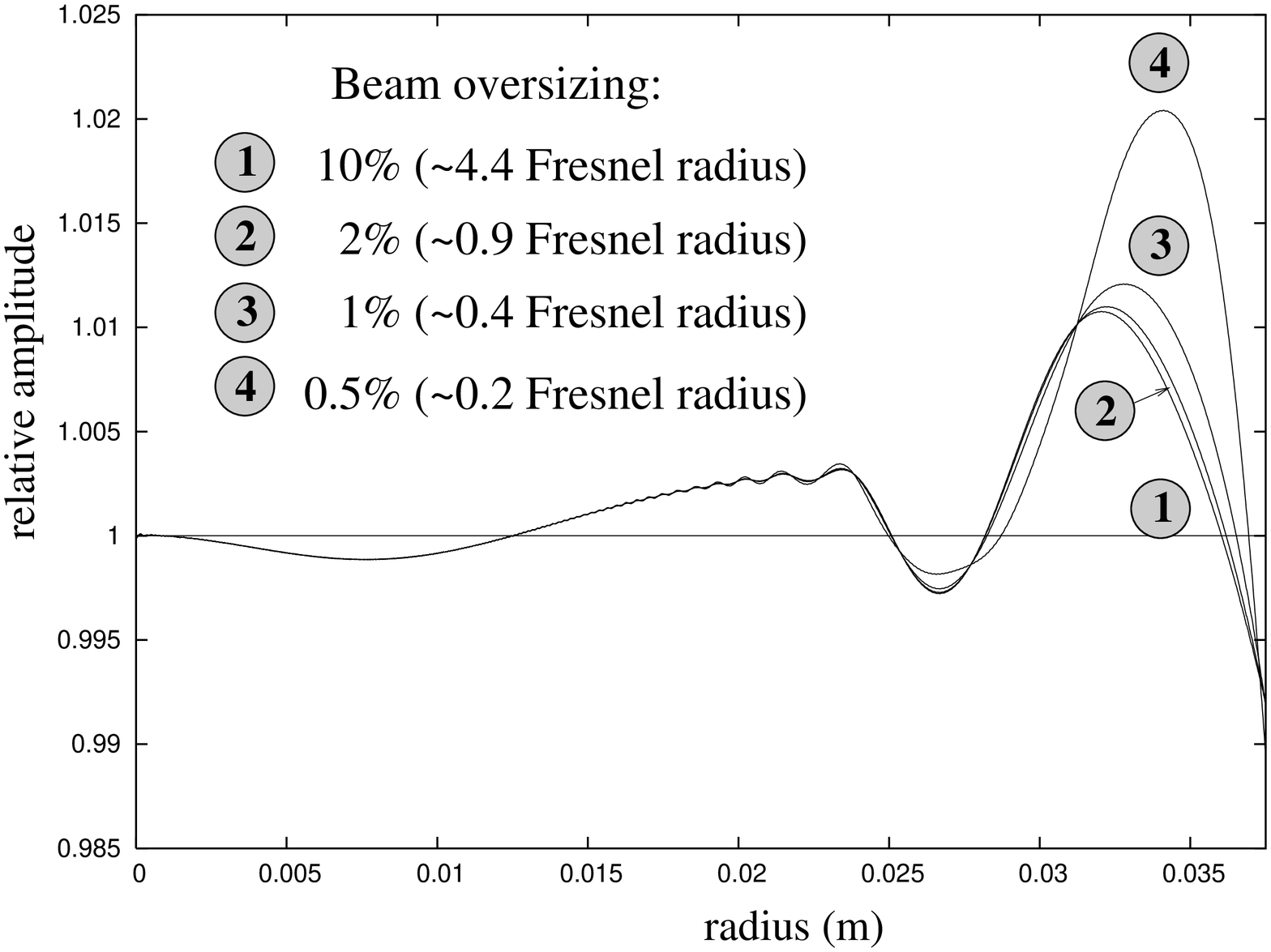,width=0.85\textwidth}
\caption{The combined LCI and boundary diffraction effects in a
"diffraction-free" PIAA apodizer ($\alpha=0.03$).
Relative amplitudes of the LCI waves and boundary wave residuals
in the output beam are shown for different ``oversizing'' widths.
}
\label{fig12}
\end{figure*}

\subsection{Toothed aperture}

The use of sawtoothlike teeth on the edge of coronagraph occulting mask
was first proposed by Purcell and Koomen (1962) to reduce flux diffracted
into the shadow region behind the mask (corresponding to the working
beam diameter here), and later proposed
to reduce diffraction errors in radiometry \cite{Boivin1978}.
The method is
based on the simple assumption that the radiation diffracted by a
straight edge
propagates in a direction perpendicular to it. According to
this simple model no  edge normal line should cross
the ``diffraction-free'' region.
For circular apertures, the  ``diffraction-free'' region
is a circle of diameter depending on
the number of teeth, the tooth depth and position
of the ``teeth'' edge relative to the ``diffraction-free'' region.
It is clear from simple geometrical considerations that the teeth should be
placed at a small distance outside the working aperture. A simple diffraction
description  of this method has been given by Shirley \& Datla (1996).
We will present our results  on this method in a following paper.

\subsection{Secondary diffraction effects in the PIAA system}

With the boundary diffraction wave now successfully reduced to
a small level, thanks to methods detailed in \S\ref{sec4_2},
the main remaining sources of diffraction effects are  regions of M1 with
strong localized curvature.
These "localized curvature induced" (LCI)
diffraction waves can be observed near the geometrical projection
of strong M1 curvature regions on
the M2 surface (corresponding to the ``transition'' where the profile
rapidly shifts from prolate to constant in Fig.~\ref{fig8}).
Compared with the boundary diffraction wave encountered in \S\ref{sec4_1}
these waves are both smoother (no small scale features)
and smaller in amplitude.
In monochromatic light, they can be corrected with a
DM (phase) and a classical apodizer (amplitude).
Such a correction, unfortunately, is not perfect for other wavelengths:
the stronger these secondary effects, the narrower we expect the bandpass
suitable for high contrast imaging to be (Fig.~\ref{fig10},~\ref{fig11}).
The regions of the maximal curvature near the boundary of the M1 mirror
are the most critical for diffraction propagation.
Two main constraints can be used to design M1's shape to minimize unwanted
LCI diffraction errors:
\begin{enumerate}
\item the maximal  curvature
of the M1 mirror should be limited to an appropriate level. This  maximal
curvature constraint also makes the manufacturing of the PIAA optics
easier.
\item the curvature should not be changing too fast and should reach
its maximal value at the boundary of the  working beam to avoid
a local curvature maximum within the working beam diameter.
In this case,
LCI diffractive waves will be
reduced due to destructive interference arising from the neighboring
areas within the working aperture diameter, and the boundary diffraction wave
can be corrected by suitable oversizing and edge apodizing  of
the beam.
\end{enumerate}

In Fig.~\ref{fig10} and~\ref{fig11}, PSFs for hybrid systems
based on amplitude Profile I and intensity profile $I_{0.1}$ are shown
in two different wavelength (0.633$\mu$m and 0.7$\mu$m). Both profiles
have an identical intensity level in the beam's ``plateau''.
The only difference is that the ``transition'' between the inner
(gaussian like) and outer (``plateau'') parts of the beam is more gentle
for
the intensity profile $I_{0.1}$. Results in  Fig.~\ref{fig10} show that
the steep transition from a prolate function in the beam to a constant
level in the beam cannot provide us with the contrast $10^{-10}$
into a wide bandpass. On the other hand,
mirror shapes corresponding to the family of much smoother intensity profiles
$I_\alpha$ (Eq.~\ref{eq3})
are good examples of PIAA optics with minimal
diffraction effects (Fig.~\ref{fig11}) and are suitable for high contrast
imaging.
We will therefore adopt these profiles
in the rest of this paper.

\section{Optimal coronagraph design and its performance}
\label{sec5}
We have shown above that the diffraction propagation effects in the PIAA
coronagraph depend on its optical design. Particularly important design
parameters are
the maximal mirror curvature
(through the parameter $\alpha$ in Eq.~\ref{eq3})
and the width $\Delta$ of the ``oversized'' part of the M1 mirror.
The coronagraph throughput, resolution and, as a result, its
performance are also determined by these parameters. In this
section we discuss how to optimize the PIAA unit design,
propose a possible solution and estimate its performance.

\subsection{Optimal choice of the beam \\ oversizing $\Delta$}
\label{sec5_1}

Our proposed ``diffraction free'' design, shown in Fig.~\ref{fig1}, gives us
a way to get
a ``diffraction free'' output beam by (1) confining
of the boundary diffraction wave outside of the working beam diameter
and (2) reducing its amplitude by edge apodization.
The cost of such a solution is to reduce the  coronagraph throughput
(by a factor equal to the relative square of the ``oversized'' area)
and resolution (by a factor $1+\Delta/D$). These losses
can be noticeable if the width $\Delta$ is large.

There are two possibilities to maintain  both a high total throughput and
high angular resolution in the PIAA system.
The first is to use the smallest value of  $\Delta$ which still
offers the required system contrast and bandpass.
Suppose that the amplitude profile can be considered
constant  next to the beam boundary, with respect to the
constant mirror curvature next to the M1 mirror edge.
Taking into account that the boundary wave is formed in the narrow
$\sim\sqrt{\lambda(z_2-z_1)}$
(width of the first Fresnel zone radius)
boundary area next to the M1 mirror edge, the width of the minimal
``oversized'' area can be as small as $\sim\sqrt{\lambda(z_2-z_1)}$.
This estimate is supported by direct diffraction calculations.
In Fig.~\ref{fig12} relative amplitudes of the LCI diffraction
waves and the boundary wave residuals, related to different ``oversizing''
widths, are presented (for $\alpha=0.03$): they
show that for $\alpha$ at least larger than 0.03 the boundary wave residuals
are negligible in comparison with the LCI diffraction wave amplitude
if $\Delta\ge\sqrt{\lambda(z_2-z_1)}$ .

\begin{table}[t]
\begin{tabular}{|c|c|c|c|}
\hline
$\alpha$ &    Resolution  ($\lambda/D$) &  Throughput &$\Delta\lambda/\lambda$ \\
\hline
0.03     &    0.96                              &  0.91             &0.074       \\
0.05     &    0.93                              &  0.86             &0.21        \\
0.1~     &    0.88                              &  0.76             &$\sim$0.6     \\
\hline
\end{tabular}
\caption{Angular resolution, total throughput and bandwith for the PIAA/CPA hybrid system
at $\lambda=0.633\mu$m. Note the total throughput is slightly larger
for a system with the edge apodizing mask affecting the working
beam.}
\label{tab1}
\end{table}
\begin{table*}[t]
\centerline{\begin{tabular}{ccc}
$r$, m               &       $P_1(r)$, m    &    $P_2(r)$, m      \\
\hline
0.000000000 & 0.000000000000000 & 1.125000000000000\\
0.000000375 & 0.000000000000021 & 1.125000000000074\\
0.000000750 & 0.000000000000087 & 1.125000000000297\\
\end{tabular}}
\caption{Shape of M1 and M2 mirrors to produce $I_{0.05}$ output beam
intensity profile. The system is designed for a 1.125~m separation
between mirrors and a collimated input/output beam.
Columns (2) and (3) include $P_1$ and $P_2$ terms (Eq.~\ref{eq2})
for each radius presented in column (1). The table is accessible
in a machine-readable format.}
\label{tab2}
\end{table*}

\begin {figure*}
\hspace{5ex}
\psfig{figure=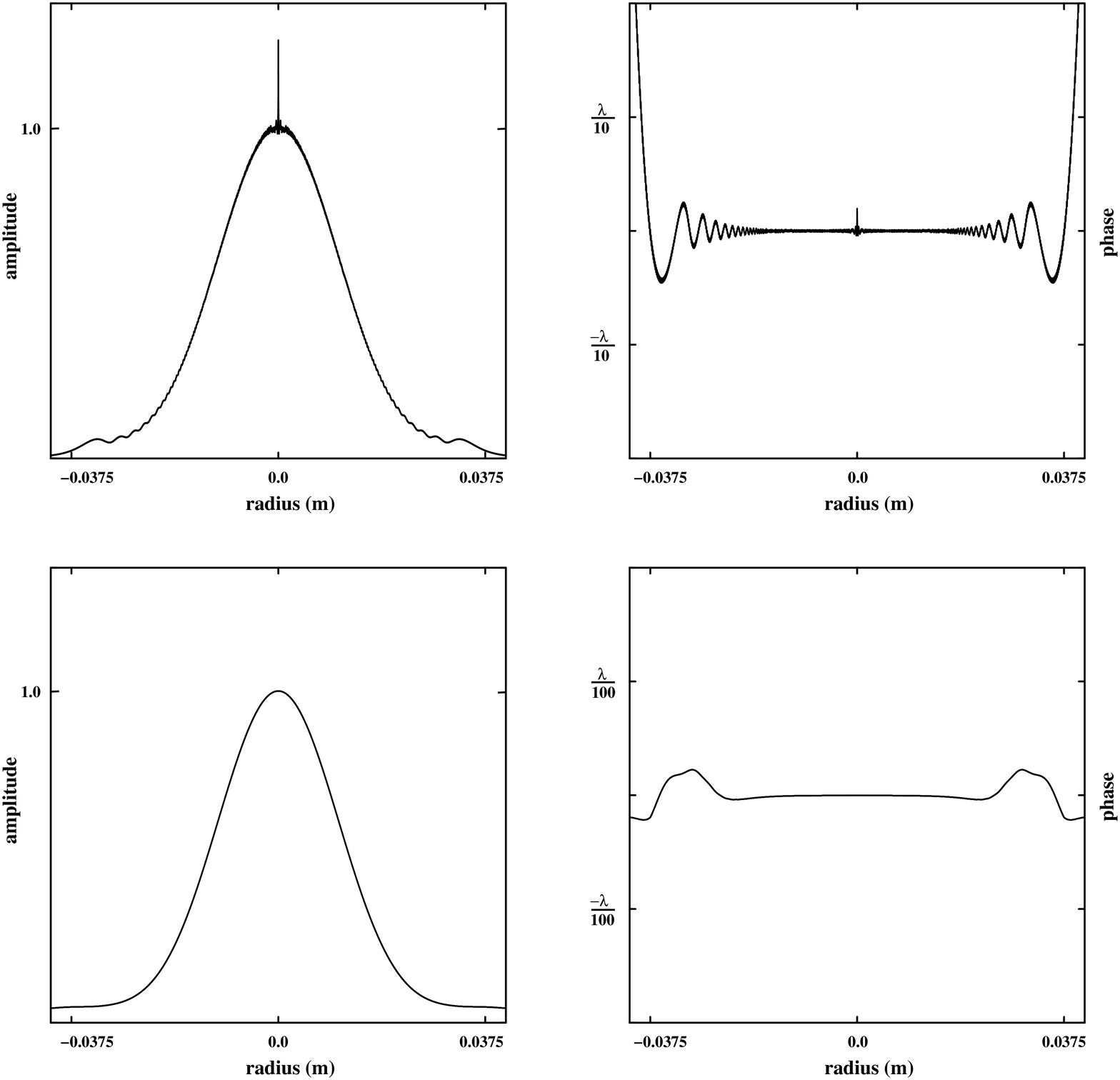,width=0.85\textwidth}
\caption{Suggested optical design: mitigation of diffraction
effects in the PIAA apodizer. Amplitude and phase distributions
on the surface of M2 mirror are shown for systems, based on the
$I_{0.05}$ intensity profile, without ({\it top})
and with ({\it bottom}) correction of diffraction effects.
}
\label{fig13}
\end{figure*}

The second possibility is to use the edge apodizing mask
(preapodizer)
partially affecting the working beam.
In this scheme, the functions of the pre-PIAA edge apodizer and the
post-PIAA classical apodizer
are shared that make it possible to slightly increase total
system throughput.

\subsection{Contrast and bandpass dependence on $\alpha$}

If the boundary wave is corrected, the system performance is determined
by the LCI diffraction waves.
For quite large values of $\alpha$ ($\sim 0.1$) the
hybrid system
based on geometric optics calculations only is sufficient to reach
a $10^{-10}$  contrast at $2\lambda/D$.
The amplitude of the LCI
waves is however higher for smaller values of  $\alpha$ (the M1 mirror
curvature increases  inside of the working beam
radius).
As a result for $\alpha\le 0.01$ the LCI waves become the
main factor affecting the PIAA coronagraph contrast.
To reach a $10^{-10}$ contrast in this case , we should use a hybrid design
which corrects not only amplitude but also phase diffraction residuals.
These diffraction residuals are highly chromatic and can not be
corrected  simultaneously
in a wide bandwidth. The bandwidth is seen to depend on $\alpha$
(Table~\ref{tab1}) and, consequently, on the system throughput.
The final hybrid PIAA design should optimize the bandwidth and the total
system throughput to reach  maximal planet detectibility.

\subsection{Suggested optical design}

We are now ready to present a preliminary design for a hybrid PIAA/CPA coronagraph
suitable for high contrast imaging of terrestrial planets.

\begin {figure*}[t]
\hspace{5ex}
\psfig{figure=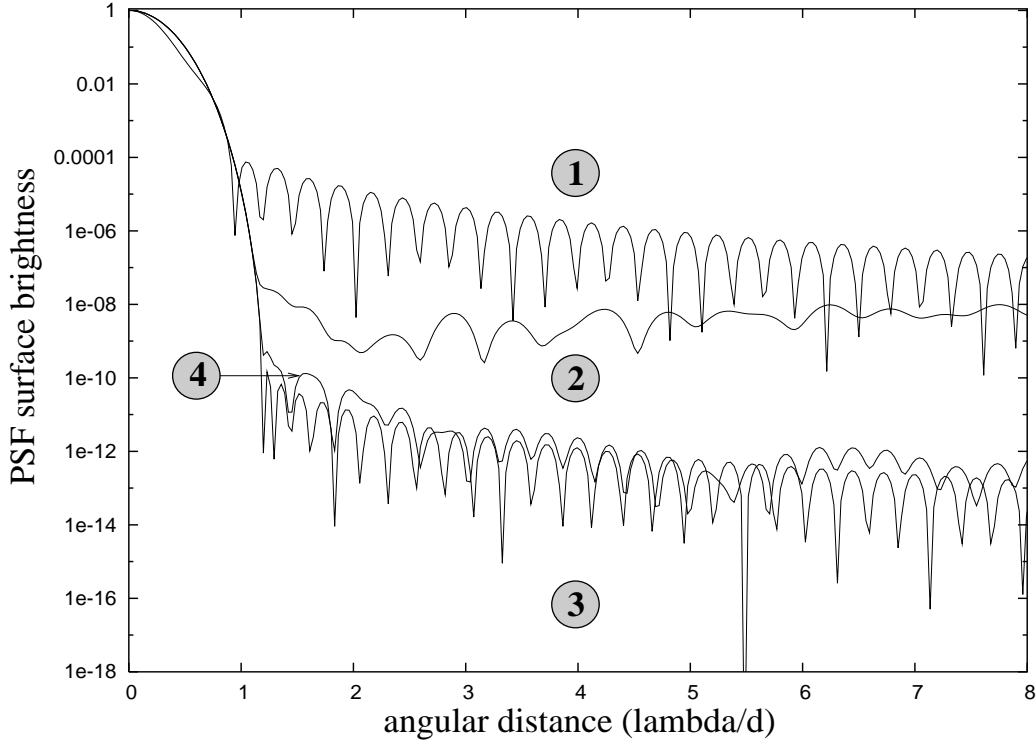,width=0.85\textwidth}
\caption{Suggested optical design: point spread functions.
The PIAA unit is built to deliver only $10^{-5}$ PSF contrast (1) by using
$I_{0.05}$ intensity profile.
Because of diffraction, the hybrid PIAA/CPA system,
(designed by assuming the geometrical optics laws) delivers only $10^{-8}$ PSF
contrast (2). The hybrid PIAA/CPA system designed
by assuming the diffraction propagation ($\lambda=0.633\mu$m) keeps
a $10^{-10}$ contrast over 21\% bandpass. The PSFs at $\lambda=0.633\mu$m
(3) and $\lambda=0.7\mu$m (4) are shown.
}
\label{fig14}
\end{figure*}

The proposed system is shown in Fig. \ref{fig1}. The shapes of both PIAA
mirrors (Fig.~\ref{fig2}, Table~\ref{tab2})
have been designed geometrically to provide us
with the $I_{0.05}$ output beam profile. This profile has been chosen
by balancing total system throughput against  bandpass
centered at 0.633~$\mu$m.
(We did not use a formal optimisation because of the cost in computer time
and because it is not needed for demonstration of feasibility.)
Although the $I_{0.05}$ output beam profile corresponds to
the contrast $10^{-5}$ at $1.5\lambda/D$ only (Fig.~\ref{fig14}),
the optics has only
5.7~cm minimal curvature radius at its edge and can be realistically
manufactured now. The additional apodization is performed by a classical
apodizer.
This apodizer removes only 10\% of the light to produce a spheroidal
prolate designed for $10^{-10}$ contrast at $1.5\lambda/D $. Taking into
account diffraction for propagation between M1 and M2 mirrors  the PSF
contrast of the geometrically designed system is limited to
$10^{-8}$ at $1.5\lambda/D$ (Fig.~\ref{fig14}).
Mitigation of the M1 edge diffraction effects (Fig. \ref{fig13}) is obtained
with
only a 2\% oversizing the entrance beam. The M1 mirror is designed to have
a constant
curvature into the oversized area, extended continuously from the working
area.
Fortunately, the mirror shape in the oversized area does not need to be
of
very high optical quality.

The oversizing width corresponds to the radius
of the first Fresnel zone ($\sqrt{\lambda(z_2-z_1)}$) for propagation
between the system mirrors and is responsible for a 4\% loss in  throughput
and 2\% loss in resolution. As a result the total throughput of the system
is equal to 86\%.
The LCI diffraction effects in the proposed system
are not more then 0.4\% and $\lambda/450$ in amplitude and phase respectively.
They are small enough
to keep the designed $10^{-10}$ contrast at $2\lambda/D$ into a 21\%
bandpass (Fig.~\ref{fig14}).

It should be noted that the
proposed design can be optimized for any designated spectral bandpass
by a formal optimisation scheme.

\subsection{Performance of the hybrid PIAAC system for direct imaging of exoplanets}

The PIAAC performance for direct detection of extrasolar planet has been
studied by \citep{Martinache2005}.
In this section, we use these results to estimate how a hybrid system would
perform, based upon differences between our original PIAAC design
\cite{Guyon2005} and the design proposed in this work.

In our hybrid design, light from the edge of the pupil is lost by the two
apodizers. The pre-remapping apodizer typically removes the outer 2\% of the
beam (4\% of its area) to avoid edge diffraction effects. The post-remapping
apodizer
typically absorbs 2\% of the light in a narrow-band system ($\alpha = 0.01$)
to 10\% of the light in a wide-band system ($\alpha = 0.05$). As shown in
Fig~\ref{fig2} , the light lost due to this second apodizer is also mostly at the
edge of the pupil.
For the purpose of performance characterization, both apodizers can be
approximated as edge-clipping masks, and we assume here that their combined
effect is to absorb 10\% of the total pupil surface (4\% from the first
apodizer, 6\% from the second apodizer), or 5\% of its radius. This estimate
is conservative, since, as explained in \S\ref{sec5_1}, the functions of
both apodizers may be combined to increase throughput.

The move from PIAAC to our hybrid design is therefore equivalent to a factor
0.93 in telescope diameter. A 4.3m diameter PIAAC hybrid telescope should
perform as well as a 4m PIAAC telescope. We recall here the main findings of
Martinache et al. (2005) with the telescope size adjusted for our new design:
\begin{itemize}
\item{In only 70~s exposure (for 100\% throughput telescope, no
zodiacal/exozodiacal light), an Earth at 10~pc would have a 50\% chance of
being detected at the SNR=5 level.}
\item{With a 4.3~m telescope, a quasi-complete detection survey of 100 F,G,K,M
type stars for ETPs can be performed in about two days of ``open shutter''
observing time (100\% throughput, 0.21 $\mu$m bandwidth centered at 0.5
$\mu$m, no zodiacal/exozodiacal light, 6 observations per star). This
observing time should be distributed in at least a year to allow for the
planets to orbit their parent stars.}
\item{With more realistic assumptions (about 10\%
telescope+coronagraph+detector throughput), a survey of 200
stars for ETPs would require about one year of observation with a 4.3~m
telescope.}
\item{Thanks to the good angular resolution of the PIAAC hybrid, the impact
of zodiacal + exozodiacal light is quite small (less than a factor of 2 in
exposure time) for systems within 10~pc with less than 2 zodi observed with a
4.3~m visible telescope. Distant systems are however more strongly affected:
the combined effect of a 2 zodi exozodiacal cloud and our zodiacal cloud is
to multiply by 3.4 the required exposure time for a face-on system at 20~pc
observed with a 4.3~m visible telescope.}
\item{Even a 2.2~m visible telescope could detect Earth-tyle planets around
a few tens of stars.}
\end{itemize}

\section{Conclusion}
\label{sec7}
Phase-induced amplitude apodization (PIAA) offers full throughput
and small IWA, but the required optics shapes are
challenging to manufacture and the technique is prone to diffraction-induced
chromatic effects. On the other hand, classical apodization coronagraphy is
very robust, but suffers from low throughput and
large IWA.
We have shown in this work that both techniques can be combined in a
``hybrid'' coronagraph design to offer high coronagraphic performance
(nearly 90\% throughput,  $1.5 \lambda/d$ IWA,
low chromaticity) with "manufacturing-friendly" optics shapes.

The system presented in \S\ref{sec5} achieves $10^{-10}$
contrast at 1.5 $\lambda/d$ and beyond in a wide spectral band
($d\lambda/\lambda \approx 0.21$) at a small cost in throughput (14\%) and
angular resolution ($7\%$). Systems with higher throughput can be designed
to operate in a smaller bandwidth.

The flexibility of our hybrid design leaves room for further optimization.
For example, the roles of the system's two apodizers can be shared to
increase throughput. Ultimately, PSF contrast, spectral bandwidth, optics
shapes and system throughput would need to be optimized for a particular
telescope size and target list.

At the $10^{-10}$ contrast level, small mirror figure errors (wether in OPD
or reflectivity) introduce chromatic aberrations in the wavefront
\cite{Shaklan2005} which require the
coronagraphic spectral bandwidth to be reduced to 10\% or less. Our study
therefore shows that a PIAA hybrid coronagraph can be designed to not be the
dominant source of chromatic aberrations.

\acknowledgements
This work was carried out under JPL contract numbers 1254445 and 1257767 for
Development of Technologies for the Terrestrial Planet Finder Mission, with
the support and hospitality of the National Astronomical Observatory of Japan.
The numerical computation has been carried on the Fujitsu PrimePower2000
supercomputer at Subaru Telescope, National Astronomical Observatory of Japan.

\end{document}